\documentclass[12pt]{article}
\usepackage{multirow,booktabs}
\usepackage{hyperref}
\usepackage{authblk}

\usepackage{xcolor}
\usepackage{amsmath}
\usepackage{graphicx}
\usepackage{epstopdf,comment}
\usepackage[numbers,compress]{natbib}
\hypersetup{colorlinks = true,
citecolor = {blue},
urlcolor = {blue},
linkcolor = {blue},
pdfstartview={FitH},
pdfstartpage={1},
pdfauthor={Marzieh Bahreman and Ming Huang and Melody Png and Bo Lan and Christopher M. Kube},
pdftitle={A pure stress formulation for modeling elastic waves using central finite differences}}

\begin{document}

\title{A pure stress formulation for modeling elastic waves using central finite differences}
\author[1]{Marzieh Bahreman}
\author[2,3]{Ming Huang}
\author[4]{Melody Png}
\author[3]{Bo Lan}
\author[1]{Christopher M. Kube\thanks{kube@psu.edu}}

\affil[1]{Department of Engineering Science and Mechanics, The Pennsylvania State University, University Park, Pennsylvania 16802, USA}
\affil[2]{Department of Engineering and Design, University of Sussex, Brighton BN1 9RH, UK}
\affil[3]{Department of Mechanical Engineering, Imperial College London, London SW7 2AZ, UK}
\affil[4]{Advanced Remanufacturing \& Technology Centre (ARTC), Agency for Science, Technology and Research (A*STAR), 3 Cleantech Loop, \#01/01 CleanTech Two, Singapore 637143, Republic of Singapore}

\date{}

\maketitle

\begin{abstract}
A pure stress-based finite difference formulation is introduced for modeling elastic wave propagation in linear elastic solids with spatial heterogeneity. The approach derives from the strong form of the elastodynamic equation of motion, in which stress is the only dependent variable. A standard second-order central difference scheme is applied to discretize the equation of motion, allowing the space-time-dependent evolution of stress components to be modeled. Numerical dispersion analysis is performed for homogeneous, elastically isotropic materials. Simulations are then carried out for a spatially heterogeneous case consisting of a bimaterial with stiffness heterogeneity. This bimaterial case allows comparison with known closed-form solutions for reflection and transmission coefficients and with an analogous displacement-based finite difference model. Simulations are executed on modern graphics processing unit architectures, enabling stress-based modeling of large-scale three-dimensional problems exceeding one billion degrees of freedom. The approach shows promise for ultrasonic simulations in materials with stiffness heterogeneity and uniform mass density, conditions common in polycrystalline metals used in engineering applications. The formulation offers a potential alternative means of modeling wave propagation and scattering in heterogeneous materials, with possible applications in nondestructive evaluation, materials characterization, biomedical ultrasound, and geosciences.
\end{abstract}

\allowdisplaybreaks

\section{Introduction}
\label{sec:I} 

Heterogeneous elastic materials arise in a large number of applications, including nondestructive testing and evaluation (NDT\&E), biomedical ultrasound, and seismic exploration. In such materials, spatial heterogeneity is derived from variations in mass density or stiffness at different locations within the volume. When elastic waves propagate through such materials, the interfaces between heterogeneities scatter and redirect portions of the wave energy. Understanding the impact of heterogeneity on elastic wave behavior is often essential, both for interpreting measured responses and for applications where the goal is to characterize the heterogeneity itself, for example, nondestructively measuring grain size in polycrystalline metals. Historically, scattering theories have been used to model the statistical properties of the wave field, such as the mean displacement for coherent or ballistic propagation~\cite{Karal_1964,Stanke_1984,Weaver_1990} or the first variation from the mean for modeling the scattered field~\cite{Weaver_1990,Ghoshal_2007}. However, modeling the actual wave field in full detail generally lies beyond the reach of analytical models, which typically require limiting assumptions such as weak scattering, frequency restrictions, or specific statistical descriptions of randomness. For these reasons, numerical methods play a key role in capturing the actual wave field evolution in heterogeneous materials. Although numerical methods can directly model elastic waves in scattering media, their application to large three-dimensional geometries at high frequencies remains a significant computational challenge. The present article addresses this by introducing a pure stress formulation as an alternative foundational starting point for numerically modeling elastic waves, with grain scattering in polycrystalline metals serving as a motivating example.

Numerical modeling of elastic wave propagation and scattering can be approached using either the strong or weak formulation. The strong formulation directly solves Cauchy's law of balance of linear momentum, whereas the weak formulation finds solutions by integrating the balance law. The choice between these two formulations has significant implications for the way in which heterogeneous material properties are handled numerically. Finite difference (FD) methods solve the strong form of elastodynamic equations directly~\cite{Temple_1988, Kelly_1976, Virieux_1986}. The most common elastodynamic equations of motion are the Navier displacement equations~\cite{Navier_1827}, which have been central to authoritative analytical works on elastic waves~\cite{Fedorov_1968, Musgrave_1970, Ben-Menahem_1981, Temple_1988, Royer_2000, Chimenti_2011}. In the strong formulation of an anisotropic Hookean solid, the displacement-based equation of motion takes the form $(C_{ijkl}u_{k,l})_{,j}= \rho\ddot{u}_i$. When the stiffness components of $C_{ijkl}$ vary spatially, the spatial derivative $(\cdot)_{,j} \equiv \partial/\partial x_j$ must also be applied to $C_{ijkl}$. In FD implementations, such derivatives must be approximated numerically~\cite{Kelly_1976}, leading to computational challenges at discontinuities where stiffness changes abruptly, for example, at grain boundaries in a polycrystal. An alternative FD approach is the velocity-stress formulation, introduced by Graves~\cite{Graves_1996} for seismic wave propagation and further developed by Moczo et al.~\cite{Moczo_2002} and others~\cite{Saenger_2004, Bohlen_2006, Kristek_2010}. The velocity-stress formulation yields a coupled system of first-order equations and avoids differentiation of the stiffness tensor. One drawback of the velocity-stress formulation in FD settings is that it requires nine field variables, three velocity components, and six stress components, making it memory-intensive for large three-dimensional problems, particularly when implemented on graphics processing units (GPUs).

Finite element (FE) methods based on the weak formulation are also commonly used to numerically model elastic waves~\cite{Pamel_2017}. The introduction of trial functions and integration over the body permits the use of Green's theorem, which eliminates spatial differentiation from the problem, including spatial derivatives on $C_{ijkl}$. FE methods are often applied in situations involving complex boundaries~\cite{Pamel_2017, Pled_2020, Kolman_2017}, whereas FD methods are generally preferred for simple rectangular domains. Recent advances in FD modeling indicate advantages in computational speed on the same order of accuracy for such domains. The choice between FD and FE methods is highly application dependent. The scope of the present article focuses on the modeling of elastic waves based on a new strong formulation, the pure stress-based equations of motion~\cite{Ignaczak_1959, Ignaczak_1963, Ostoja-Starzewski_2019, Kube_2021, Kube_2022}. As these pure stress-based equations are being employed in a numerical context for the first time, we adopt FD methods and, for simplicity, second-order central differences to demonstrate their numerical solutions in three dimensions for homogeneous and bimaterial configurations.

The remainder of the paper is organized as follows. Sec.~\ref{sec:II} presents the theoretical framework for stress-based equations of motion, contrasting them with displacement-based and velocity-stress approaches. Sec.~\ref{sec:III} details the implementation of FD on GPUs, including discretization schemes and boundary conditions. Sec.~\ref{sec:IV} is divided into two parts: the first addresses numerical dispersion for different mesh configurations, and the second examines a model problem consisting of two half-spaces of gold and tungsten to evaluate reflection and transmission characteristics at their interface. This example allows for comparison of the FD results with analytical solutions and a commercially available FE software package. Finally, Sec.~\ref{sec:V} discusses the findings, computational advantages, and future directions.

\section{Elastodynamic Formulations for Wave Propagation}
\label{sec:II} 
\subsection{Displacement, velocity-stress, and stress-based equations of motion}
\label{sec:II:A}

 This section presents a complete stress-based formulation, accounting for heterogeneity in both mass density and stiffness, to establish a thorough theoretical foundation. Consider a linear elastic material with spatially varying mass density $\rho(\mathbf{x})$ and spatially varying anisotropic stiffness $C_{ijkl}(\mathbf{x})$. The balance of linear momentum is expressed through Cauchy's law, 
\begin{align}
\sigma_{ij,j} &= \rho\ddot{u}_i,
\label{Eq1}
\end{align}
which relates the stress gradient ($\boldsymbol{\sigma}$) to the acceleration of the wave displacement ($\mathbf{u}$). Throughout, index notation is used to express components of vectors and higher-order tensors, and the commas denote spatial differentiation. Often, Hooke's law ($\sigma_{ij} = C_{ijkl}\varepsilon_{kl} = C_{ijkl}u_{k,l}$) is substituted into Eq.~(\ref{Eq1}) to arrive at the displacement-based equation of motion,
\begin{align}
(C_{ijkl}u_{k,l})_{,j} &= \rho\ddot u_i,
\label{Eq2}
\end{align}
where spatially dependent quantities are implied for brevity. If the spatial heterogeneity in stiffness vanishes, Eq.~(\ref{Eq2}) reduces to the standard second-order wave equation,
\begin{align}
C_{ijkl}u_{k,jl} &= \rho\ddot u_i.
\label{Eq3}
\end{align}
Assuming a fixed Cartesian ($x, y, z$) coordinate system, we express Eq.~(\ref{Eq2}) in explicit component form, where the indices $i, j, k, l$ correspond to the $x, y, z$ directions. For notational convenience, Voigt notation is adopted to represent the elastic stiffness tensor, contracting $C_{ijkl}$ to its two-index equivalent $C_{mn}$ $(m, n = 1,\ldots,6)$. This yields three coupled equations:
\begin{align}
\rho\ddot{u}_x &= \partial_x(C_{11}\partial_x u_x+C_{12}\partial_y u_y + C_{13}\partial_z u_z \notag\\
&\qquad+ C_{14}(\partial_yu_z+\partial_zu_y)+C_{15}(\partial_xu_z+\partial_zu_x)+C_{16}(\partial_xu_y+\partial_yu_x))\notag\\
&\quad+\partial_z(C_{15}\partial_xu_x+C_{25}\partial_yu_y+C_{35}\partial_zu_z \notag\\
&\qquad+C_{45}(\partial_yu_z+\partial_zu_y)+C_{55}(\partial_xu_z+\partial_zu_x)+C_{56}(\partial_xu_y+\partial_yu_x))\notag\\
&\quad+\partial_y(C_{16}\partial_xu_x+C_{26}\partial_yu_y+C_{36}\partial_zu_z \notag\\
&\qquad+C_{46}(\partial_yu_z+\partial_zu_y)+C_{56}(\partial_xu_z+\partial_zu_x)+C_{66}(\partial_xu_y+\partial_yu_x)),
\label{Eq4}\\
\rho\ddot{u}_y &= \partial_y(C_{12}\partial_xu_x+C_{22}\partial_yu_y+C_{23}\partial_zu_z \notag\\
&\qquad+C_{24}(\partial_yu_z+\partial_zu_y)+C_{25}(\partial_xu_z+\partial_zu_x)+C_{26}(\partial_xu_y+\partial_yu_x))\notag\\
&\quad+\partial_z(C_{14}\partial_xu_x+C_{24}\partial_yu_y+C_{34}\partial_zu_z \notag\\
&\qquad+C_{44}(\partial_yu_z+\partial_zu_y)+C_{45}(\partial_xu_z+\partial_zu_x)+C_{46}(\partial_xu_y+\partial_yu_x))\notag\\
&\quad+\partial_x(C_{16}\partial_x u_x+C_{26}\partial_y u_y + C_{36}\partial_z u_z \notag\\
&\qquad+C_{46}(\partial_yu_z+\partial_zu_y)+C_{56}(\partial_xu_z+\partial_zu_x)+C_{66}(\partial_xu_y+\partial_yu_x)),
\label{Eq5}\\
\rho\ddot{u}_z &= \partial_z(C_{13}\partial_xu_x+C_{23}\partial_yu_y+C_{33}\partial_zu_z \notag\\
&\qquad+C_{34}(\partial_yu_z+\partial_zu_y)+C_{35}(\partial_xu_z+\partial_zu_x)+C_{36}(\partial_xu_y+\partial_yu_x))\notag\\
&\quad+\partial_y(C_{14}\partial_xu_x+C_{24}\partial_yu_y+C_{34}\partial_zu_z \notag\\
&\qquad+C_{44}(\partial_yu_z+\partial_zu_y)+C_{45}(\partial_xu_z+\partial_zu_x)+C_{46}(\partial_xu_y+\partial_yu_x))\notag\\
&\quad+\partial_x(C_{15}\partial_x u_x+C_{25}\partial_y u_y + C_{35}\partial_z u_z \notag\\
&\qquad+C_{45}(\partial_yu_z+\partial_zu_y)+C_{55}(\partial_xu_z+\partial_zu_x)+C_{56}(\partial_xu_y+\partial_yu_x)),
\label{Eq6}
\end{align}
where $\partial_x$, $\partial_y$, and $\partial_z$ represent spatial derivatives with respect to $x$, $y$, and $z$, respectively. In Eqs.~(\ref{Eq4})--(\ref{Eq6}), all 21 stiffness components are operated on by a spatial derivative. Using Eqs.~(\ref{Eq4})--(\ref{Eq6}) within an FD scheme requires all spatial derivatives to be approximated numerically, meaning that discontinuities associated with spatial heterogeneity can be problematic if not handled with care. Additionally, terms involving mixed derivatives, such as $\partial_x[C_{12}\partial_y(u_y)]$ do not have standard FD representations. Kelly et al.~\cite{Kelly_1976} addressed this by using the average value of $C_{12}$ between neighboring cells before applying a central difference approximation to the mixed derivative. Similarly, Temple~\cite{Temple_1988} adopted the approach of Kelly et al.\ for FD simulations in three-dimensional heterogeneous materials. It is worth noting that spatial heterogeneity in mass density $\rho$ does not present differentiation issues, since $\rho$ appears in Eqs.~(\ref{Eq4})--(\ref{Eq6}) without spatial derivatives.

An alternative to the pure displacement-based equations are the so-called velocity-stress equations, which are most commonly employed in modern FD schemes. The velocity-stress equations consist of a system of nine first-order partial differential equations: three governing the velocity ($\mathbf{v}$) components,
\begin{align}
\rho\dot{v}_x &= \partial_x\sigma_x+\partial_y\sigma_{xy}+\partial_z\sigma_{xz},
\label{Eq7}\\
\rho\dot{v}_y &= \partial_x\sigma_{xy}+\partial_y\sigma_{y}+\partial_z\sigma_{yz},
\label{Eq8} \\
\rho\dot{v}_z &= \partial_x\sigma_{xz}+\partial_y\sigma_{yz}+\partial_z\sigma_{z},
\label{Eq9}
\end{align}
which are obtained by substituting $\ddot{\mathbf{u}} = \dot{\mathbf{v}}$ into the right-hand side of Eq.~(\ref{Eq1}) and expanding $\sigma_{ij,j}$. The remaining six equations govern the stress components, obtained by differentiating Hooke's law with respect to time, $\dot{\sigma}_{ij} = C_{ijkl}\dot{u}_{k,l} = C_{ijkl}v_{k,l}$, which leads to
\begin{align}
\dot{\sigma}_{x} &= C_{11}\partial_xv_x+C_{12}\partial_yv_y+C_{13}\partial_zv_z \notag\\
&\quad+C_{14}(\partial_yv_z+\partial_zv_y)+C_{15}(\partial_xv_z+\partial_zv_x)+C_{16}(\partial_xv_y+\partial_yv_x),
\label{Eq10}\\
\dot{\sigma}_{y} &= C_{12}\partial_xv_x+C_{22}\partial_yv_y+C_{23}\partial_zv_z \notag\\
&\quad+C_{24}(\partial_yv_z+\partial_zv_y)+C_{25}(\partial_xv_z+\partial_zv_x)+C_{26}(\partial_xv_y+\partial_yv_x),
\label{Eq11}\\
\dot{\sigma}_{z} &= C_{13}\partial_xv_x+C_{23}\partial_yv_y+C_{33}\partial_zv_z \notag\\
&\quad+C_{34}(\partial_yv_z+\partial_zv_y)+C_{35}(\partial_xv_z+\partial_zv_x)+C_{36}(\partial_xv_y+\partial_yv_x),
\label{Eq12}\\
\dot{\sigma}_{yz} &= C_{14}\partial_xv_x+C_{24}\partial_yv_y+C_{34}\partial_zv_z \notag\\
&\quad+C_{44}(\partial_yv_z+\partial_zv_y)+C_{45}(\partial_xv_z+\partial_zv_x)+C_{46}(\partial_xv_y+\partial_yv_x),
\label{Eq13}\\
\dot{\sigma}_{xz} &= C_{15}\partial_xv_x+C_{25}\partial_yv_y+C_{35}\partial_zv_z \notag\\
&\quad+C_{45}(\partial_yv_z+\partial_zv_y)+C_{55}(\partial_xv_z+\partial_zv_x)+C_{56}(\partial_xv_y+\partial_yv_x),
\label{Eq14}\\
\dot{\sigma}_{xy} &= C_{16}\partial_xv_x+C_{26}\partial_yv_y+C_{36}\partial_zv_z \notag\\
&\quad+C_{46}(\partial_yv_z+\partial_zv_y)+C_{56}(\partial_xv_z+\partial_zv_x)+C_{66}(\partial_xv_y+\partial_yv_x).
\label{Eq15}
\end{align}

The velocity-stress equations involve only simple first-order derivatives, which are well-suited for numerical approximation. Furthermore, they do not present the differentiation issues associated with stiffness components that arise in the displacement-based formulation, as discussed previously.

Here, a strong formulation based on equations of motion containing only stress components as dependent variables is derived. To the best of our knowledge, this is the first instance in which the pure stress equations of motion are cast into a strong form suitable for numerical modeling of elastic waves.
 
We begin by considering the two equivalent forms of Eq.~(\ref{Eq1}), 
\begin{align}
\sigma_{ik,k} &= \rho\ddot u_i,
\label{Eq16}\\
\sigma_{jk,k} &= \rho\ddot u_j.
\label{Eq17} 
\end{align}
Taking the spatial derivative of Eq.~(\ref{Eq16}) with respect to $x_j$, the spatial derivative of Eq.~(\ref{Eq17}) with respect to $x_i$, and summing the two results leads to
\begin{align}
\sigma_{ik,jk}+\sigma_{jk,ik} &= \rho_{,i}\ddot u_j+\rho_{,j}\ddot u_i+\rho(\ddot u_{i,j}+\ddot u_{j,i}),
\label{Eq18}
\end{align}
where the product rule was applied to the $\rho\mathbf{u}$ terms. Noting that the infinitesimal strain tensor is $\varepsilon_{ij} = \frac{1}{2}(u_{i,j} + u_{j,i})$, Eq.~(\ref{Eq18}) can be rewritten as
\begin{align}
\frac{1}{2}(\sigma_{ik,jk}+\sigma_{jk,ik}) &= \frac{1}{2}(\rho_{,i}\ddot u_j+\rho_{,j}\ddot u_i)+\rho\ddot\varepsilon_{ij}.
\label{Eq19}
\end{align}
Strain is related to stress through the inverse of Hooke's law $\varepsilon_{ij} = S_{ijkl}\sigma_{kl}$, where $S_{ijkl}$ is the compliance tensor. The compliance and stiffness tensors are tensor inverses of one another, $C_{ijmn}S_{mnkl}=I_{ijkl}=(\delta_{ik}\delta_{jl}+\delta_{il}\delta_{jk})/2$. Thus, Eq.~(\ref{Eq19}) can be rewritten as
\begin{align}
\frac{1}{2}(\sigma_{ik,jk}+\sigma_{jk,ik}) &= I_{ijkl}\rho_{,k}\ddot u_l+\rho S_{ijkl}\ddot\sigma_{kl}.
\label{Eq20} 
\end{align}
From Eq.~(\ref{Eq1}), we have $\ddot{u}_l = \sigma_{lm,m}/\rho$; thus, Eq.~(\ref{Eq20}) becomes
\begin{align}
\frac{1}{2}(\sigma_{ik,jk}+\sigma_{jk,ik}) &= I_{ijkl}\sigma_{lm,m}\frac{\rho_{,k}}{\rho}+\rho S_{ijkl}\ddot\sigma_{kl}.
\label{Eq21}
\end{align}
Relabeling $i \to p,\; j \to q$ and then taking the inner product of Eq.~(\ref{Eq21}) with $C_{ijpq}$, and noting that $C_{ijpq}S_{pqkl} = I_{ijkl}$ and $C_{ijpq}I_{pqmn} = C_{ijmn}$, and after further relabeling of the dummy indices to restore the standard $k$ and $l$ notation, we arrive at the final form
\begin{align}
\ddot\sigma_{ij} &= \rho^{-1}C_{ijkl}\sigma_{km,ml}+C_{ijkl}\sigma_{km,m}(\rho^{-1})_{,l}.
\label{Eq22}
\end{align}
Taking into account the symmetry of the stress tensor, the six governing equations represented by Eq.~(\ref{Eq22}) can be reformulated to express each stress component in explicit form. In what follows, the Einstein summation convention is replaced by explicit summation over the Cartesian components $x$, $y$, and $z$. For notational convenience, the abbreviated Voigt-like notation $\sigma_x \equiv \sigma_{xx}$, $\sigma_y \equiv \sigma_{yy}$, and $\sigma_z \equiv \sigma_{zz}$ is adopted for the normal stress components, and the Voigt notation is adopted for the stiffness tensor, contracting $C_{ijkl}$ to its two-index equivalent $C_{mn}$ $(m, n = 1,\ldots,6)$. Vectors and matrices are denoted using bracket notation, with subscripts indicating dimensions: $[\cdot]_{6\times1}$ denotes a six-component column vector and $[\cdot]_{6\times6}$ denotes a $6\times6$ matrix. The stress evolution equations can then be written as

\begin{align}
\begin{bmatrix}
\ddot{\sigma}_{x} \\
\ddot{\sigma}_{y} \\
\ddot{\sigma}_{z}\\
\ddot{\sigma}_{yz} \\
\ddot{\sigma}_{xz} \\
\ddot{\sigma}_{xy}
\end{bmatrix} = 
\begin{bmatrix}
C_{11} & C_{12} & C_{13} & C_{14} & C_{15} & C_{16} \\
C_{12} & C_{22} & C_{23} & C_{24} & C_{25} & C_{26} \\
C_{13} & C_{23} & C_{33} & C_{34} & C_{35} & C_{36} \\
C_{14} & C_{24} & C_{34} & C_{44} & C_{45} & C_{46} \\
C_{15} & C_{25} & C_{35} & C_{45} & C_{55} & C_{56} \\
C_{16} & C_{26} & C_{36} & C_{46} & C_{56} & C_{66}
\end{bmatrix}
\left(\begin{bmatrix}
K_1 \\
K_2 \\
K_3 \\
K_4 \\
K_5 \\
K_6
\end{bmatrix}+\begin{bmatrix}
M_1 \\
M_2 \\
M_3 \\
M_4 \\
M_5 \\
M_6
\end{bmatrix}\right),
\label{Eq23}
\end{align}
where
\begin{align}
K_1 &= \rho^{-1}\partial_{x}(\partial_{x}\sigma_x+\partial_{y}\sigma_{xy}+\partial_{z}\sigma_{xz}),\notag\\
K_2 &= \rho^{-1}\partial_{y}(\partial_{x}\sigma_{xy}+\partial_{y}\sigma_{y}+\partial_{z}\sigma_{yz}),\notag\\
K_3 &= \rho^{-1}\partial_{z}(\partial_{x}\sigma_{xz}+\partial_{y}\sigma_{yz}+\partial_{z}\sigma_{z}),\notag\\
K_4 &= \rho^{-1}[\partial_{y}(\partial_{x}\sigma_{xz}+\partial_{y}\sigma_{yz}+\partial_{z}\sigma_{z})+\partial_{z}(\partial_{x}\sigma_{xy}+\partial_{y}\sigma_{y}+\partial_{z}\sigma_{yz})],\notag\\
K_5 &= \rho^{-1}[\partial_{x}(\partial_{x}\sigma_{xz}+\partial_{y}\sigma_{yz}+\partial_{z}\sigma_{z})+\partial_{z}(\partial_{x}\sigma_{x}+\partial_{y}\sigma_{xy}+\partial_{z}\sigma_{xz})],\notag\\
K_6 &= \rho^{-1}[\partial_{x}(\partial_{x}\sigma_{xy}+\partial_{y}\sigma_{y}+\partial_{z}\sigma_{yz})+\partial_{y}(\partial_{x}\sigma_{x}+\partial_{y}\sigma_{xy}+\partial_{z}\sigma_{xz})],\notag\\
M_1 &=(\partial_{x}\sigma_x+\partial_{y}\sigma_{xy}+\partial_{z}\sigma_{xz})\partial_{x}(\rho^{-1}),\notag\\
M_2 &=(\partial_{x}\sigma_{xy}+\partial_{y}\sigma_{y}+\partial_{z}\sigma_{yz})\partial_y(\rho^{-1}),\notag\\
M_3 &=(\partial_{x}\sigma_{xz}+\partial_{y}\sigma_{yz}+\partial_{z}\sigma_{z})\partial_z(\rho^{-1}),\notag\\
M_4 &=(\partial_{x}\sigma_{xz}+\partial_{y}\sigma_{yz}+\partial_{z}\sigma_{z})\partial_y(\rho^{-1})+(\partial_{x}\sigma_{xy}+\partial_{y}\sigma_{y}+\partial_{z}\sigma_{yz})\partial_z(\rho^{-1}),\notag\\
M_5 &=(\partial_{x}\sigma_{xz}+\partial_{y}\sigma_{yz}+\partial_{z}\sigma_{z})\partial_x(\rho^{-1})+(\partial_{x}\sigma_{x}+\partial_{y}\sigma_{xy}+\partial_{z}\sigma_{xz})\partial_z(\rho^{-1}),\notag\\
M_6 &=(\partial_{x}\sigma_{xy}+\partial_{y}\sigma_{y}+\partial_{z}\sigma_{yz})\partial_x(\rho^{-1})+(\partial_{x}\sigma_{x}+\partial_{y}\sigma_{xy}+\partial_{z}\sigma_{xz})\partial_y(\rho^{-1}).
\label{Eq24}
\end{align}
It is worth noting that $\rho\ddot{u}_x = \partial_{x}\sigma_x + \partial_{y}\sigma_{xy} + \partial_{z}\sigma_{xz}$, $\rho\ddot{u}_y = \partial_{x}\sigma_{xy} + \partial_{y}\sigma_{y} + \partial_{z}\sigma_{yz}$, and $\rho\ddot{u}_z = \partial_{x}\sigma_{xz} + \partial_{y}\sigma_{yz} + \partial_{z}\sigma_{z}$, which shows how $\ddot{\boldsymbol{\sigma}}$ can be related back to displacements or velocities ($\dot{\mathbf{v}} = \ddot{\mathbf{u}}$). For example, examining a term such as $K_1 - M_1$ more closely, we observe via the chain rule that
\begin{align}
K_1+M_1 &= [\rho^{-1}\partial_{x}+\partial_{x}(\rho^{-1})](\partial_{x}\sigma_x
+\partial_{y}\sigma_{xy}+\partial_{z}\sigma_{xz})\notag\\
&= [\rho^{-1}\partial_{x}+\partial_{x}(\rho^{-1})](\rho\ddot{u}_x)\notag\\
&= \rho^{-1}\partial_{x}(\rho\ddot{u}_x)+\rho\ddot{u}_x\partial_{x}(\rho^{-1})\notag\\
&= \partial_x(\ddot{u}_x).
\label{Eq25}
\end{align}
A similar reduction holds for the remaining terms, which allows Eq.~(\ref{Eq23}) to be rewritten as
\begin{align}
\begin{bmatrix}
\ddot{\sigma}_{x} \\
\ddot{\sigma}_{y} \\
\ddot{\sigma}_{z}\\
\ddot{\sigma}_{yz} \\
\ddot{\sigma}_{xz} \\
\ddot{\sigma}_{xy}
\end{bmatrix} = 
\begin{bmatrix}
C_{11} & C_{12} & C_{13} & C_{14} & C_{15} & C_{16} \\
C_{12} & C_{22} & C_{23} & C_{24} & C_{25} & C_{26} \\
C_{13} & C_{23} & C_{33} & C_{34} & C_{35} & C_{36} \\
C_{14} & C_{24} & C_{34} & C_{44} & C_{45} & C_{46} \\
C_{15} & C_{25} & C_{35} & C_{45} & C_{55} & C_{56} \\
C_{16} & C_{26} & C_{36} & C_{46} & C_{56} & C_{66}
\end{bmatrix}
\begin{bmatrix}
\partial_x(\ddot{u}_x) \\
\partial_y(\ddot{u}_y) \\
\partial_z(\ddot{u}_z) \\
\partial_y(\ddot{u}_z)+\partial_z(\ddot{u}_y) \\
\partial_x(\ddot{u}_z)+\partial_z(\ddot{u}_x) \\
\partial_x(\ddot{u}_y)+\partial_y(\ddot{u}_x)
\end{bmatrix},
\label{Eq26}
\end{align}
which is consistent with the acceleration observed via Hooke's law, $\ddot{\sigma}_{ij} = C_{ijkl}\ddot{u}_{k,l}$, as expected. A related consistency check involves the potential energy, as the energy function must remain consistent regardless of whether it is expressed in terms of displacement, strain, or stress. The standard potential energy for a linear elastic Hookean solid expressed in terms of strain, sometimes referred to as strain energy, is
\begin{align}
\rho\psi &= \frac{1}{2}C_{ijkl}\varepsilon_{ij}\varepsilon_{kl}=\frac{1}{2}C_{ijkl}u_{i,j}u_{k,l},
\label{Eq27}
\end{align}
which follows from a Taylor series expansion about a zero-strain reference state. Noting that $C_{ijmn}S_{mnkl} = I_{ijkl}$ and substituting $\varepsilon_{ij} = S_{ijkl}\sigma_{kl}$ into Eq.~(\ref{Eq27}), it can be shown that 
\begin{align}
\rho\psi &= \frac{1}{2}S_{ijkl}\sigma_{ij}\sigma_{kl}.\label{Eq28}
\end{align}
It should be noted that Eq.~(\ref{Eq28}) is commonly referred to as the complementary energy. As a side note, not part of the present linear elastic framework, in nonlinear elasticity, the complementary energy is denoted $\psi^*$ and satisfies $\psi + \psi^* = \sigma_{ij}\varepsilon_{ij}$, where $\sigma$ and $\varepsilon$ are the second Piola--Kirchhoff stress and Green strain, respectively. The assumption of linear elasticity yields $\psi = \psi^*$, confirming that the strain energy and complementary energy coincide in the present setting. For wave simulations, it is sometimes desirable to visualize wave evolution through this unambiguous scalar energy quantity, which can be computed from the six stress components at each time step.

In this section, we have recapitulated the equations of motion formulated in terms of displacement and velocity-stress variables, and introduced the pure stress equations of motion given in Eqs.~(\ref{Eq23})--(\ref{Eq24}). The pure stress equations and their application within an FD framework are the focus of the present article. Before proceeding to the numerical implementation, the following section provides additional context relating the three forms of the equations of motion.

\subsection{Discussion on equations of motion for finite difference implementations}
\label{sec:II:B}

The stress equations presented in Eqs.~(\ref{Eq23})--(\ref{Eq24}) provide an alternative formulation for FD simulations of elastic waves in spatially heterogeneous materials. It is important to clarify that the advantages of stress-based formulations discussed in this section apply specifically to strong-form numerical methods such as FD schemes. Weak formulation methods, particularly FE approaches, do not require spatial derivatives of material properties regardless of whether displacement or stress variables are used, as the integration by parts inherent in weak formulations eliminates this requirement. Therefore, the computational benefits of stress-based formulations are most relevant for FD and other strong-form discretization methods.

Before proceeding with discretization of the stress equations, we briefly contrast the three forms of the equations of motion within the context of FD implementations and highlight the specific benefits of using the pure stress equations for FD simulations. From a computational memory perspective, the displacement equations require storage of three components, $u_x$, $u_y$, and $u_z$; the velocity-stress equations require nine components, $v_x$, $v_y$, $v_z$, $\sigma_x$, $\sigma_y$, $\sigma_z$, $\sigma_{yz}$, $\sigma_{xz}$, and $\sigma_{xy}$; while the stress equations require only six stress components, $\sigma_x$, $\sigma_y$, $\sigma_z$, $\sigma_{yz}$, $\sigma_{xz}$, and $\sigma_{xy}$. For each case, all components must be stored over the entire three-dimensional grid, along with neighboring values depending on the FD approximation used. Consequently, the displacement equations require the least memory, followed by the pure stress equations, and then the velocity-stress equations. However, the velocity-stress equations involve only first-order derivatives, compared to second-order derivatives in the displacement and stress formulations, and therefore require fewer arithmetic operations per time step. Thus, a trade-off exists between memory usage and computational cost, the balance of which is application dependent. In recent years, many scattering problems have required spatial discretization of heterogeneities at scales much smaller than the acoustic wavelength; for example, simulations of ultrasonic waves in polycrystals. In such cases, memory becomes the primary constraint, often preventing practical three-dimensional simulations of experimentally relevant configurations.

Another consideration is how boundary conditions are handled for each formulation of the equations of motion. For a medium with heterogeneities within the volume, all three approaches implicitly satisfy the conditions at the internal boundaries and interfaces. However, for the displacement equations in a strong form, the spatial derivatives of the stiffness $C_{ijkl}$ appear explicitly, as seen in Eqs.~(\ref{Eq4})--(\ref{Eq6}). In FD simulations based on strong formulations, these derivatives must be approximated numerically using finite differences~\cite{Kelly_1976}. This contrasts with weak formulation methods (e.g., FE), where integration by parts eliminates the need for derivatives of material properties. Consequently, internal boundaries separating regions of varying stiffness must be approximated in displacement-based FD schemes, and the accuracy of this approximation diminishes as the stiffness contrast increases. The stress equations in Eqs.~(\ref{Eq23})--(\ref{Eq24}) contain spatial derivatives of the density rather than the stiffness. Thus, similar approximations are needed to handle density variations across interfaces separating heterogeneities of differing density. The stress equations offer an advantage here, as derivative approximations are required only for the scalar density field rather than all 21 components of the stiffness tensor. Additionally, the appearance of the logarithmic derivative is expected to improve accuracy, as the natural logarithm acts to smooth variations across an interface. In cases where the mass density is spatially uniform, as is often the case for polycrystalline metals, no derivative approximations on density are needed, making the stress formulation particularly attractive. Finally, it is noted that the velocity-stress equations require no derivative approximations on either the density or stiffness components.

Boundary conditions on the exterior surface of the domain require separate consideration. A common boundary condition is the prescription of a traction (sometimes referred to as a traction force),
\begin{align}
\tau_i &= \sigma_{ij}n_j,
\label{Eq29}
\end{align}
where $\mathbf{n}$ is the unit vector normal to the surface of the solid domain. The prescribed traction depends on the particular application. For example, traction-free conditions ($\sigma_{ij}n_j = 0$) are commonly prescribed on free surfaces, as is typical in nondestructive evaluation applications at solid--air interfaces. When the pressure $P$ in a surrounding fluid must be taken into account or when a source of excitation is applied, the traction is prescribed as $\tau_i = P n_i$. The classification of boundary conditions as Dirichlet or Neumann depends on which field variable is treated as the primary unknown. In the stress-based formulation, the stress components are the dependent variables; therefore, directly prescribing the traction $\tau_i = \sigma_{ij}n_j$ constitutes a Dirichlet boundary condition, as it specifies the value of the primary unknown on the boundary. In contrast, imposing the same physical traction on the displacement-based equations of motion requires
\begin{align}
\tau_i = \sigma_{ij}n_j = C_{ijkl}u_{k,l}n_j,
\label{Eq30}
\end{align}
which represents a Neumann boundary condition, as it prescribes a derivative of the primary unknown (displacement) rather than its value directly. Thus, the stress-based formulation naturally accommodates the traction prescription as a Dirichlet boundary condition, avoiding the additional numerical complexity associated with Neumann conditions in displacement-based FD schemes.

The foregoing discussion of computational considerations and boundary conditions illustrates the potential advantages of using stress equations within FD simulations. Furthermore, FD simulations are generally simpler to implement and more tractable than FE schemes. The Dirichlet boundary conditions inherent in the stress-based formulation further support the simplicity, tractability, and accessibility of FD solvers. Thus, the remainder of the article focuses on the FD implementation of the stress equations.

\section{Finite Difference Scheme and Problem Configuration}
\label{sec:III} 
\subsection{Discretization of the stress-based equations}
\label{sec:III:A} 

To implement the stress-based equations of motion numerically, we employ a central difference FD scheme, which discretizes both the spatial and temporal domains to permit computational solutions to the wave propagation problem. Let $a$, $b$, and $c$ denote the mesh spacing in the $x$, $y$, and $z$ directions, respectively. For spatial derivatives with respect to the same coordinate direction, 
\begin{align}
\frac{\partial^2 f}{\partial \xi^2} &\approx \frac{f(\xi+h) - 2f(\xi) + f(\xi-h)}{h^2},
\label{Eq31}
\end{align}
where $\xi \in \{x, y, z\}$ and $h \in \{a, b, c\}$. Similarly, for mixed spatial derivatives,
\begin{align}
\frac{\partial^2 f}{\partial \xi \partial \eta} &\approx \frac{f(\xi+h_1,\eta+h_2) - f(\xi+h_1,\eta-h_2)}{4h_1h_2} \notag\\
&\quad - \frac{f(\xi-h_1,\eta+h_2) - f(\xi-h_1,\eta-h_2)}{4h_1h_2},
\label{Eq32}
\end{align}
where $\xi, \eta \in \{x, y, z\}$ with $\xi \neq \eta$, and $h_1$, $h_2 \in \{a, b, c\}$ are the corresponding mesh spacings for $\xi$ and $\eta$, respectively. The second-order temporal derivatives are approximated as
\begin{align}
\frac{\partial^2 f}{\partial t^2} &\approx \frac{f(t+\Delta t) - 2f(t) + f(t-\Delta t)}{(\Delta t)^2},
\label{Eq33}
\end{align}
where $\Delta t$ is the time step. In all cases, $f$ represents any of the six stress components ($\sigma_x$, $\sigma_y$, $\sigma_z$, $\sigma_{yz}$, $\sigma_{xz}$, $\sigma_{xy}$) in the spatial and temporal discretization. For the first-order derivatives of the mass density $\rho$, we have
\begin{align}
\frac{\partial(\rho^{-1})}{\partial\xi} &\approx \frac{\rho^{-1}(\xi+h)-\rho^{-1}(\xi-h)}{2h}.
\label{Eq34}
\end{align}
With these approximations, a time-marching matrix equation based on Eq.~(\ref{Eq23}) can be formed, 
\begin{align}
\left[\boldsymbol{\sigma}_{\text{6x1}}(x,y,z,t+\Delta t)\right] &= 2\left[\boldsymbol{\sigma}_{\text{6x1}}(x,y,z,t)\right] 
-\left[\boldsymbol{\sigma}_{\text{6x1}}(x,y,z,t-\Delta t)\right] \notag\\
&\quad+(\Delta t)^2\left[\textbf{C}_{\text{6x6}}\right]
\left(\left[\textbf{K}_{\text{6x1}}\right]+\left[\textbf{M}_{\text{6x1}}\right]\right),
\label{Eq35}
\end{align}
where the approximated spatial derivatives are embedded within the vectors $[\mathbf{K}_{6\times 1}]$ and $[\mathbf{M}_{6\times 1}]$. Equation~(\ref{Eq35}) is the pure stress FD scheme employed in the following sections.

A notable finding is that the pure stress equations contain spatial derivatives of the scalar mass density rather than those of the fourth-rank elastic stiffness tensor, as appears in the displacement equations of Eq.~(\ref{Eq2}). This is particularly advantageous when studying polycrystalline materials in which the mass density is commonly assumed to be uniform. In such cases, the vector $[\mathbf{M}_{6\times 1}]$, which contains the derivative approximations of $\rho$, vanishes in Eq.~(\ref{Eq35}), yielding
\begin{align}
\left[\boldsymbol{\sigma}_{\text{6x1}}(x,y,z,t+\Delta t)\right] &= 2\left[\boldsymbol{\sigma}_{\text{6x1}}(x,y,z,t)\right] 
-\left[\boldsymbol{\sigma}_{\text{6x1}}(x,y,z,t-\Delta t)\right] \notag\\
&\quad+(\Delta t)^2\left[\textbf{C}_{\text{6x6}}\right]\left[\textbf{K}_{\text{6x1}}\right],
\label{Eq36}
\end{align}
when implemented in the FD scheme, or $\ddot{\sigma}_{ij} = \rho^{-1}C_{ijkl}\sigma_{km,lm}$ in full form. The compact form of Eq.~(\ref{Eq36}), its straightforward implementation in FD, and its particular utility for studying elastic wave scattering in polycrystalline materials motivated the results presented in Sec.~\ref{sec:IV} to focus on materials with uniform mass density.

\subsection{Simulation details}
\label{sec:III:B} 

\begin{figure}[t]
\centering
\includegraphics[width=0.8\textwidth]{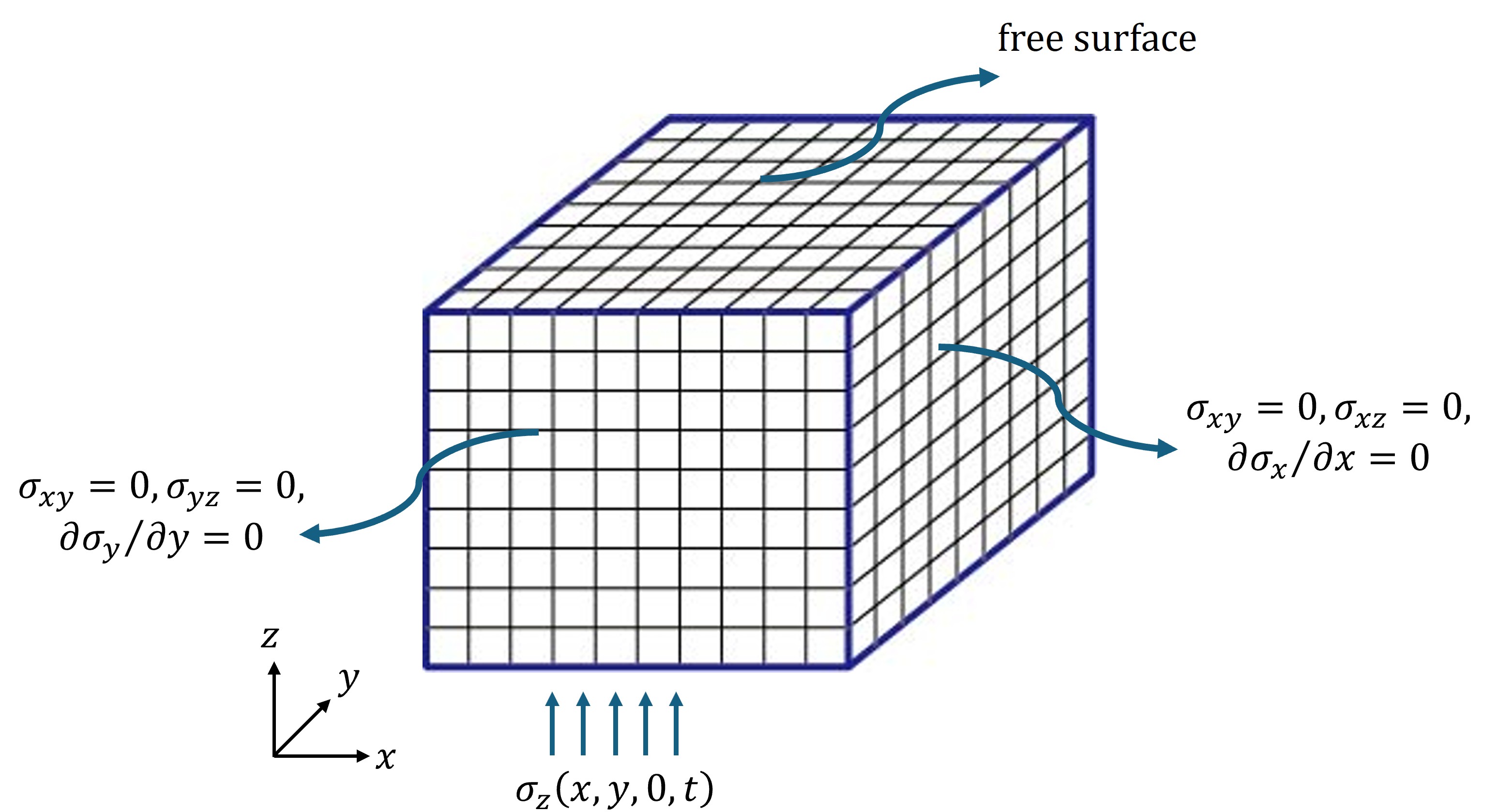}
\caption{Schematic view of the three-dimensional computational domain with dimensions $L_x \times L_y \times L_z$, showing the coordinate system, initial stress excitation at $z=0$, lateral boundary conditions with zero shear stresses and zero normal stress gradients ($\partial\sigma/\partial n = 0$) consistent with the translational symmetry of plane wave propagation in the lateral directions, and traction-free top surface.}
\label{fig:FIG1}
\end{figure}

Sec.~\ref{sec:III:A} described how the stress equations are discretized, including the central difference approximations to the derivatives leading to Eq.~(\ref{Eq36}). In this section, the remaining simulation details used to produce the results in Sec.~\ref{sec:IV} are provided. Figure~\ref{fig:FIG1} illustrates the initial and boundary conditions applied to the three-dimensional domain. The computational domain is defined as $0 \leq x \leq L_x$, $0 \leq y \leq L_y$, and $0 \leq z \leq L_z$; within the FD scheme, boundary conditions are enforced by assigning appropriate values to the stress components at boundary nodes and at ghost nodes located outside the computational domain. The wave is initialized on the face defined by the plane $(x, y, z = 0)$ using a temporally modulated sine wave pulse centered at $t_0$,
\begin{align}
\sigma_z(x,y,0,t) &= -A\sin{(\omega_0 (t-t_0))}\exp{\left(\frac{-(t-t_0)^2}{2\Sigma_t^2}\right)},
\label{Eq37}
\end{align}
where $A$ is the stress amplitude, $\omega_0 = 2\pi f_0$ is the angular frequency at the center frequency $f_0$, and $\Sigma_t$ is the Gaussian width parameter. The choice of a constant amplitude $A$ in Eq.~(\ref{Eq37}) approximates plane wave propagation, which is the focus of the results in Sec.~\ref{sec:IV}. Furthermore, the boundary conditions are selected to maintain plane wave propagation in the $z$ direction and to prevent spurious reflections. With $(x, y, z = 0)$ as the excitation plane, the four lateral faces are $(x = 0, y, z)$, $(x = L_x, y, z)$, $(x, y = 0, z)$, and $(x, y = L_y, z)$. The boundary conditions on these lateral faces are set such that shear stresses vanish,
\begin{align}
\sigma_{xy} &= 0, \sigma_{xz}=0 \quad \text{at} \quad (x=0,y,z) \text{ and } (x=L_x,y,z), \notag\\
\sigma_{xy} &= 0, \sigma_{yz}=0 \quad \text{at} \quad (x,y=0,z), \text{ and }(x,y=L_y,z).
\label{Eq38}
\end{align}
These conditions are consistent with plane wave propagation in the $z$-direction, for which the stress field is uniform in $x$ and $y$. In this case, all shear stress components that coupling the propagation direction to the lateral directions vanish identically throughout the domain. For the normal stress components, the plane wave solution exhibits no lateral variation, and accordingly a zero normal gradient condition $\partial\sigma/\partial n = 0$ is enforced on these faces. This is implemented by setting ghost node values equal to their corresponding interior boundary node values, which provides the necessary values for FD derivative calculations at the lateral boundaries. Together, these conditions reflect the translational symmetry of the plane wave field in the lateral directions, effectively rendering the lateral boundaries transparent to the propagating wave.

The top surface at $(x, y, z = L_z)$ is assumed to be a free surface with a traction-free boundary condition,
\begin{align}
\sigma_{z} &= 0, \sigma_{xz}=0, \sigma_{yz}=0 \quad \text{at} \quad z=L_z.
\label{Eq39}
\end{align}

The stress-based FD method is implemented on GPUs to handle the computationally intensive nature of the three-dimensional simulations. Two NVIDIA RTX A6000 ADA Edition GPUs were used, each with 49.85~GB of memory and 1457~TFLOPS of computational throughput.

The simulations are designed to model elastic wave propagation in a homogeneous isotropic aluminum sample with realistic material properties and wave characteristics. We begin with single-crystal aluminum elastic constants in Voigt notation: $C_{11} = 103.4~\text{GPa}$, $C_{12} = 57.1~\text{GPa}$, $C_{44} = 28.6~\text{GPa}$, with density $\rho = 2700~\text{kg/m}^3$. An averaging procedure is then applied to obtain equivalent isotropic constants. The anisotropy factor is computed as $\nu = C_{11} - C_{12} - 2C_{44} = -10.9~\text{GPa}$ and redistributed equally among the elastic constants: $C_{12} = 54.92~\text{GPa}$, $C_{44} = 26.42~\text{GPa}$, and $C_{11} = C_{12} + 2C_{44} = 107.76~\text{GPa}$. For the isotropic case, $C_{22} = C_{33} = C_{11}$, $C_{13} = C_{23} = C_{12}$, and $C_{55} = C_{66} = C_{44}$, while all shear-normal coupling terms ($C_{14}$, $C_{15}$, $C_{16}$, etc.) are set to zero. Three models are considered in the following discussion, with their spatial and temporal characteristics listed in Table~\ref{tab:TABLEI}. The computational parameters were determined as follows. First, the mesh spacing $c$ was specified for each model ($0.023$, $0.028$, and $0.034$~mm), which, together with fixed grid dimensions ($N_x \times N_y \times N_z = 401 \times 401 \times 1201$), determined the physical domain size via $L_x = L_y = (N_x - 1) \times a$ and $L_z = (N_z - 1) \times c$, where $a = b = c$ for uniform mesh spacing. The degrees of freedom were calculated as $\text{d.o.f.} = N_x \times N_y \times N_z \times 6$, accounting for the six independent stress components. The longitudinal wave velocity was computed as $V_0 = \sqrt{C_{11}/\rho} = \sqrt{107.76 \times 10^9/2700} = 6317.5~\text{m/s}$. To ensure numerical stability, the Courant--Friedrichs--Lewy (CFL) condition is enforced. For the present three-dimensional stress-based formulation, this condition takes the form
\begin{align}
\Delta t &\leq \frac{C h}{V_{max}},
\label{Eq40}
\end{align}
where $C$ is the Courant number, $h = \min(a, b, c)$ is the smallest mesh spacing, and $V_{\max}$ is the maximum wave velocity in the medium. The time step $\Delta t$ is thus determined by the mesh resolution, wave speed, and stability requirements. Using $C = 0.3$ and $V_{\max} = V_0$, the time step was calculated as $\Delta t = 0.3 \times c/V_0$. The total simulation time was estimated as $t \approx L_z/V_0$ to allow the wave to traverse the entire domain, and the number of time steps was calculated as $N_t = t/\Delta t$. This approach ensures that as the mesh spacing increases across the three models, the time step increases correspondingly to maintain numerical stability, while the number of time steps decreases, keeping the total simulation time proportional to the domain length.

These values confirm that all three models provide sufficient spatial sampling to accurately capture wave propagation at 5~MHz, with the finest mesh (Model~1) yielding the highest resolution of approximately 55 points per wavelength. The domain sizes are chosen to span sufficient wavelengths in all directions: approximately 7--11 wavelengths in the lateral directions ($x$ and $y$) and 22--32 wavelengths in the propagation direction ($z$). This ensures an adequate domain size to capture the complete wave evolution, while minimizing spurious boundary reflections within the simulation time window.
\begin{table}[ht]
\caption{Different homogeneous models. Grid points $N_x \times N_y \times N_z$, dimensions $L_x \times L_y \times L_z$ ($mm^3$), mesh spacing $c$ ($mm$), degrees of freedom (dof), points per wavelength (ppw), wavelengths per dimension $N_{\lambda x} = N_{\lambda y}$ and $N_{\lambda z}$, time step $\Delta t$ ($ns$), number of time steps $N_t$, and total simulation time $t$ ($\mu s$).}
\resizebox{\textwidth}{!}{
\begin{tabular}{c c c c c c c c c c}
\hline\hline
Model ($N_x \times N_y \times N_z$) & $L_x \times L_y \times L_z$ & $c$ & dof & ppw & $N_{\lambda x,y}$ & $N_{\lambda z}$ & $\Delta t$ & $N_t$ & $t$ \\
\hline
$401 \times 401 \times 1201$ & $9.38 \times 9.38 \times 28.14$ & $0.023$ & $1.16 \times 10^9$ & $\sim$55 & $\sim$7 & $\sim$22 & $1.11$ & $5,810$ & $6.45$ \\
$401 \times 401 \times 1201$ & $11.32 \times 11.32 \times 33.96$ & $0.028$ & $1.16 \times 10^9$ & $\sim$45 & $\sim$9 & $\sim$27 & $1.34$ & $5,507$ & $7.37$ \\
$401 \times 401 \times 1201$ & $13.66 \times 13.66 \times 40.99$ & $0.034$ & $1.16 \times 10^9$ & $\sim$37 & $\sim$11 & $\sim$32 & $1.62$ & $5,234$ & $8.48$ \\
\hline\hline
\end{tabular}
}
\label{tab:TABLEI}
\end{table}

\begin{figure}[t]
\centering
\includegraphics[width=0.8\textwidth]{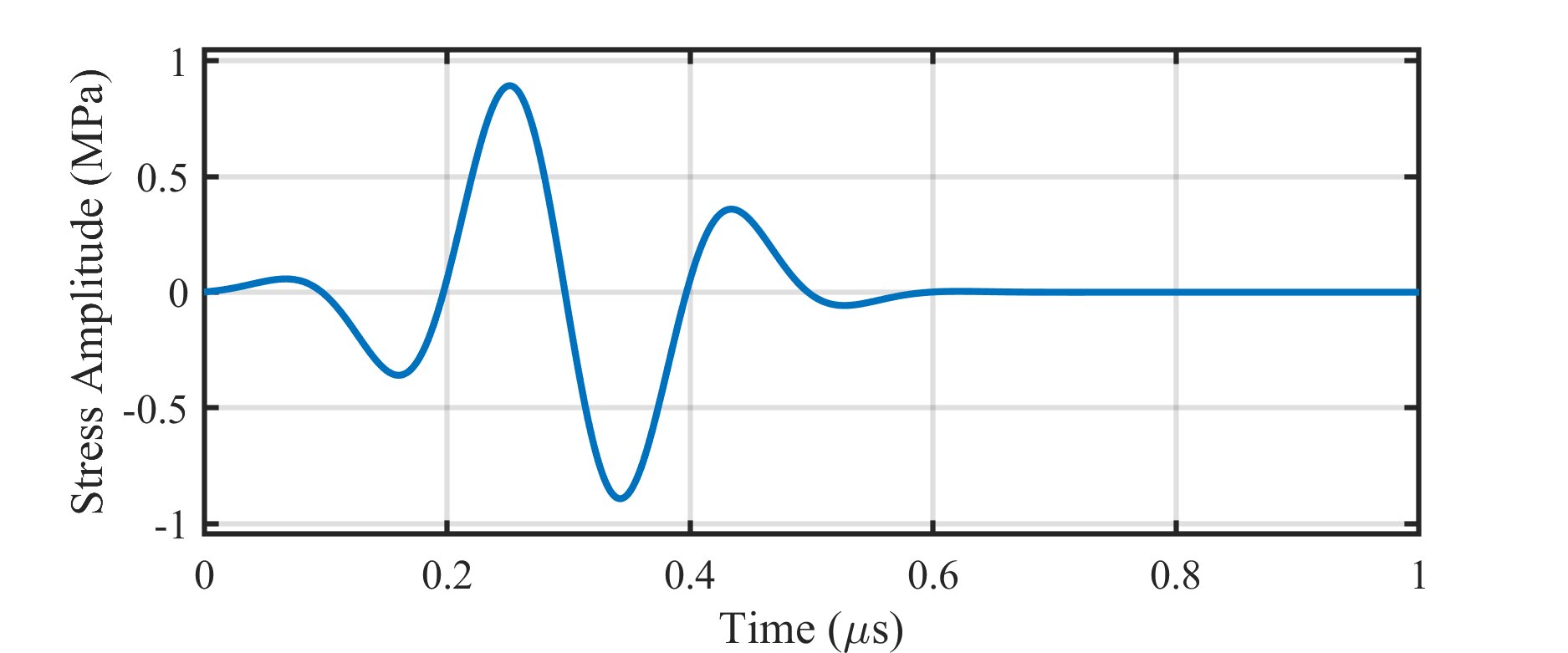}
\caption{Temporal profile of the initial Gaussian-modulated sinusoidal pulse with center frequency of 5 MHz and pulse width of 0.1 $\mu$s, applied as the stress boundary condition at $z=0$.}
\label{fig:FIG2}
\end{figure}

The initial excitation is a longitudinal wave with $A = 1$~MPa, center frequency $f_0 = 5$~MHz, and pulse width $\Sigma_t = 0.1$~$\mu$s. The temporal profile of the initial pulse is shown in Figure~\ref{fig:FIG2}, which corresponds to the Gaussian-modulated pulse described by Eq.~(\ref{Eq37}). Excitation is introduced through a stress boundary condition at $z = 0$, applied uniformly across the $x$--$y$ plane.

\section{Results}
\label{sec:IV}
\subsection{Dispersion analysis}
\label{sec:IV:A}

An essential aspect of validating the stress-based FD method is evaluating its numerical dispersion characteristics. Numerical dispersion, inherent in discrete approximations of wave equations, can lead to phase and amplitude errors that accumulate over time and space. In this section, we examine the numerical phase and amplitude errors in relation to analytical predictions.

\subsubsection{Analytical prediction of numerical errors}
\label{sec:IV:A:1}
To assess numerical errors in the stress-based FD method, we derive an analytical dispersion relation for a monochromatic plane wave in a homogeneous isotropic medium~\cite{Huang_2020}. The exact solution for a plane wave propagating in an isotropic material is
\begin{align}
w(z,t) &= p_0e^{-i(kz-2\pi ft)},
\label{Eq41}
\end{align}
where $p_0$ is the wave amplitude, $k$ is the wavenumber, and $f$ is the frequency.

The stress--strain relationship $\sigma_{ij} = C_{ijkl}u_{k,l}$ is expressed in explicit component form with respect to a fixed Cartesian coordinate system $(x, y, z)$, where the displacement components $u_x$, $u_y$, $u_z$ are denoted $u$, $v$, $w$, respectively, and the Voigt notation is adopted for the stiffness tensor, contracting $C_{ijkl}$ to its two-index equivalent $C_{mn}$ $(m, n = 1,\ldots,6)$. For an isotropic material, the resulting stress components are the following:
\begin{align}
\sigma_x &= C_{11}u_{,x}+C_{12}v_{,y}+C_{13}w_{,z}, \notag\\
\sigma_y &= C_{12}u_{,x}+C_{22}v_{,y}+C_{23}w_{,z}, \notag\\
\sigma_z &= C_{13}u_{,x}+C_{23}v_{,y}+C_{33}w_{,z}, \notag\\
\sigma_{yz} &= C_{44}(v_{,z}+w_{,y}), \notag\\
\sigma_{xz} &= C_{55}(u_{,z}+w_{,x}), \notag\\
\sigma_{xy} &= C_{66}(u_{,y}+v_{,x}).
\label{Eq42}
\end{align}
For $z$-direction propagation with $C_{13} = C_{23} = C_{12}$ and $C_{33} = C_{11}$, Eq.~(\ref{Eq42}) reduces to
\begin{align}
\sigma_x &= \sigma_y=C_{12}w_{,z}, \sigma_z=C_{11}w_{,z}, \notag\\
\sigma_{xy} &= 0, \sigma_{xz}=0, \sigma_{yz}=0.
\label{Eq43}
\end{align}
Substituting Eq.~(\ref{Eq41}) into Eq.~(\ref{Eq43}) yields
\begin{align}
\sigma_x &= \sigma_y=-iC_{12}p_0ke^{-i(kz-2\pi ft)} \notag\\
\sigma_z &= -iC_{11}p_0ke^{-i(kz-2\pi ft)}.
\label{Eq44}
\end{align}
Substituting these into the FD equations gives the dispersion relation
\begin{align}
\cos{(2\pi f\Delta t)}-1 &= \frac{C_{11}(\Delta t)^2}{\rho c^2}
[\cos{(kc)}-1],
\label{Eq45} 
\end{align}
where $c$ is the mesh spacing in the $z$ direction. Solving for $k$ yields
\begin{align}
k &= \frac{1}{c}\arccos{\{\frac{\cos{(2\pi f\Delta t)}-1}{V_0^2(\Delta t)^2}c^2+1\}},
\label{Eq46} 
\end{align}
where $V_0$ is the longitudinal wave velocity.

The theoretical phase error is
\begin{align}
\delta_{\phi_{Theoretical}} &= \frac{k}{k_0}-1=\frac{S}{2\pi}\arccos{\{\frac{\cos{(2\pi C/S)-1}}{C^2}+1\}}-1,
\label{Eq47}
\end{align}
where $S = V_0/(fc)$ is the spatial sampling number and $C = V_0\Delta t/c$ is the Courant number.

\subsubsection{Evaluation of numerical phase errors}
\label{sec:IV:A:2}

To evaluate numerical phase errors in the stress-based FD method, simulations were performed using the model parameters listed in Table~\ref{tab:TABLEI}, comparing the numerically obtained phase differences with analytical predictions in various spatial sampling numbers ($S$) and Courant numbers ($C$).

The simulation setup consisted of a three-dimensional aluminum model with characteristics listed in Table~\ref{tab:TABLEI}. The model was excited with a pulse of the center frequency $5$~MHz, and stress signals were recorded at two $z$ positions (approximately $1/4$ and $3/4$ of the height of the model). The numerical phase difference was calculated by: (1) windowing the recorded signals to isolate the main pulse, (2) performing a fast Fourier transform (FFT) on the windowed signals, and (3) computing the phase difference between the two signals.

\begin{figure}[t]
\centering
\includegraphics[width=0.8\textwidth]{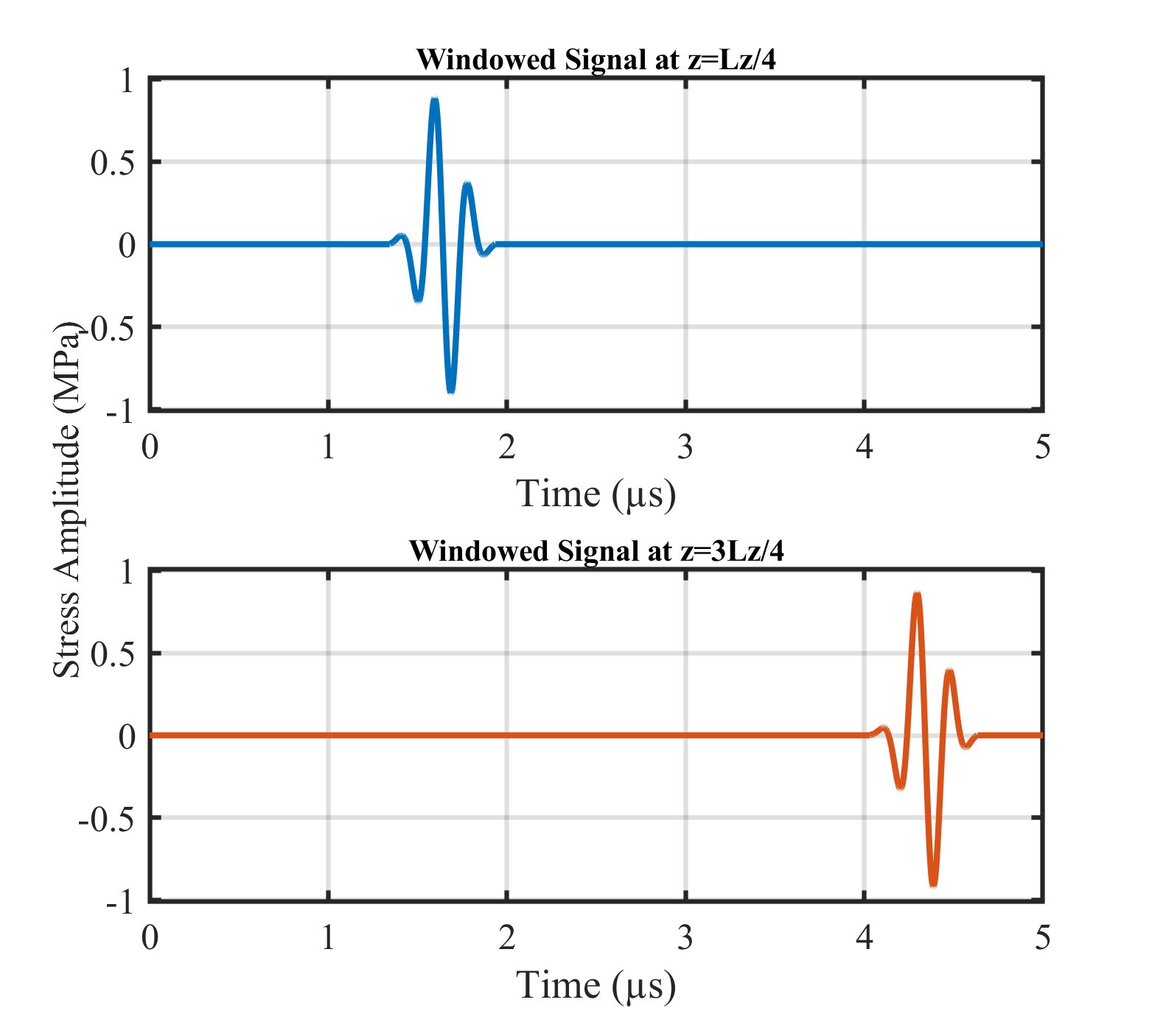}
\caption{Windowed stress signals recorded at $z=L_z/4$ (top) and $z=3L_z/4$ (bottom) in the aluminum specimen after excitation with a 5 MHz pulse, showing stable wave propagation with maintained waveform characteristics used for phase difference analysis.}
\label{fig:FIG3}
\end{figure}

Figure~\ref{fig:FIG3} shows the stress waves recorded at the two positions for the second model in Table~\ref{tab:TABLEI}, using $C = 0.3$. Both signals maintain similar waveform characteristics, demonstrating stable wave propagation. The windowing process eliminates boundary reflections and secondary wave modes, ensuring that the phase difference calculations reflect the propagation of the primary wave mode.

The numerical phase error was calculated from the phase difference between the two stress signals ($\phi$) as
\begin{align}
\delta_{\phi_{Numerical}} &= \frac{\phi}{2\pi f dz/V_0}-1,
\label{Eq48}
\end{align}
where $dz$ is the distance between the measurement points and $V_0 = \sqrt{C_{11}/\rho} = 6{,}317.5$~m/s is the longitudinal wave velocity. The relative phase error $\Delta\phi$ is
\begin{align}
\Delta\phi &= \lvert\frac{\delta_{\phi_{Numerical}}-\delta_{\phi_{Theoretical}}}{\delta_{\phi_{Theoretical}}}\rvert\times 100\%.
\label{Eq49} 
\end{align}

\begin{figure}[t]
\centering
\includegraphics[width=0.8\textwidth]{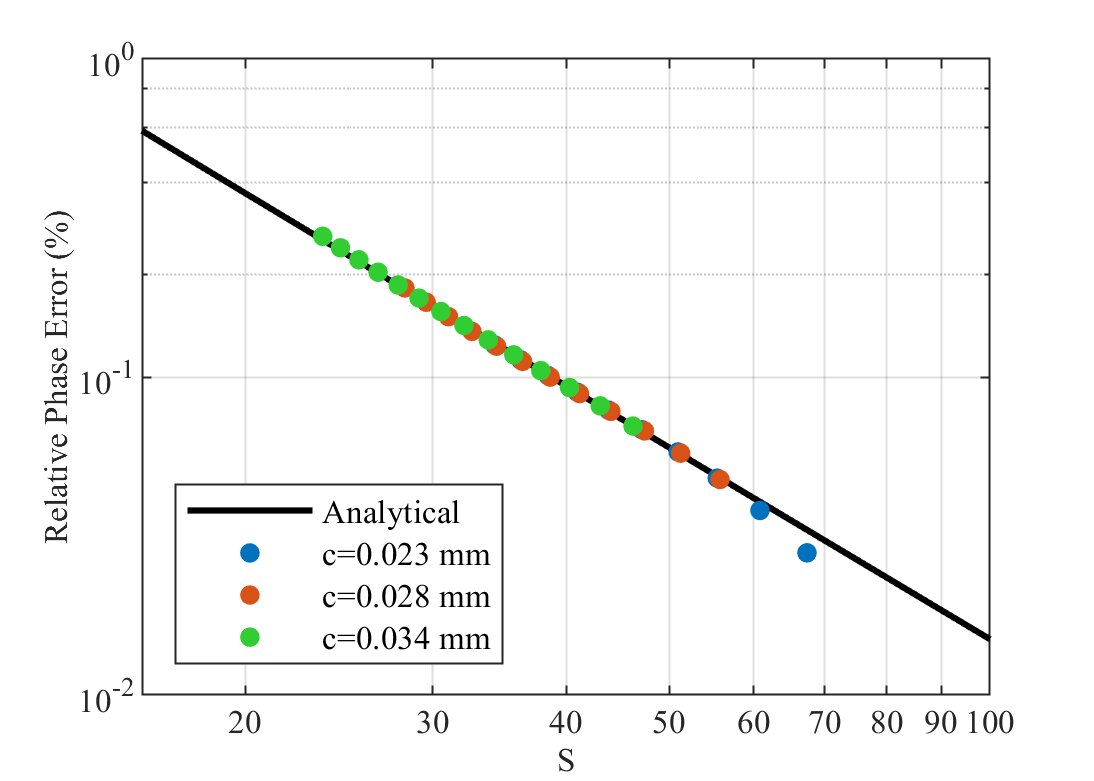}
\caption{Numerical phase errors as a function of spatial sampling number $S$ for three mesh sizes (0.023, 0.028, and 0.034 mm) at Courant number $C=0.3$. The solid black line shows analytical predictions, while colored dots represent numerical results.}
\label{fig:FIG4}
\end{figure}

To investigate the relationship between spatial sampling number $S$ and phase error, simulations were performed with varying mesh spacing over a frequency range of $4$--$8$~MHz. Figure~\ref{fig:FIG4} presents numerical phase errors versus $S$ for three mesh sizes ($c = 0.023$, $0.028$, and $0.034$~mm) compared to analytical predictions (solid black line), with $C = 0.3$. The results demonstrate excellent agreement between the simulations and analytical predictions in all spatial sampling numbers.

The relative phase error decreases from approximately $0.29\%$ at $S = 23.12$ to $0.028\%$ at $S = 67.33$. The mesh $0.028$~mm  shows the best agreement with analytical predictions, effectively capturing wave physics while minimizing discretization effects. Comparison with FE results from Huang~\cite{Huang_2020} shows that both methods exhibit decreasing phase errors with increasing $S$. The FE results span $S > 10$ with errors beginning at approximately $0.5\%$, while the present FD results span $S > 23$ with initial errors of $0.29\%$, reflecting the different mesh sizes and frequency ranges used in each study.

\begin{figure}[t]
\centering
\includegraphics[width=0.8\textwidth]{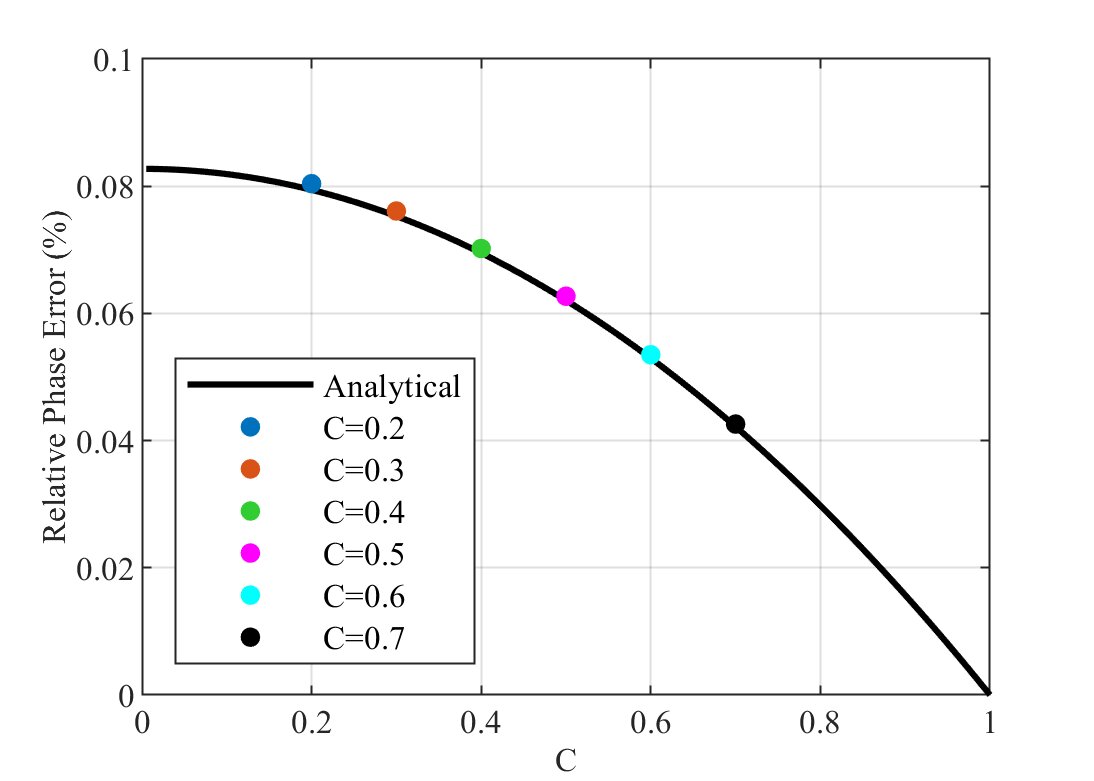}
\caption{Numerical phase errors as a function of Courant number $C$ for fixed mesh size of 0.028 mm. The solid black line shows analytical predictions, while colored dots represent numerical results for $C$ values ranging from 0.2 to 0.7.}
\label{fig:FIG5}
\end{figure}

Figure~\ref{fig:FIG5} illustrates the relationship between numerical phase errors and the Courant number for $C = 0.2$ to $0.7$ using a fixed mesh size of $c = 0.028$~mm. Relative phase errors decrease nonlinearly from approximately $0.08\%$ at $C = 0.2$ to $0.043\%$ at $C = 0.7$, with excellent agreement between numerical results (colored dots) and analytical predictions (solid black line). This behavior arises because, as $C$ approaches unity, the temporal and spatial discretization errors partially cancel each other, reducing the net phase error. At the limit $C = 1$, phase error can be nearly eliminated; however, this value is not recommended in practice, as numerical dispersion can locally accelerate wave velocities beyond the theoretical maximum, effectively violating the stability condition, and the limit $C = 1$ cannot be simultaneously achieved for all wave modes in materials with spatial property variations. Compared to the FE study of Huang~\cite{Huang_2020}, which used $c = 0.025$~mm and $C = 0.4$--$1.0$ at $5$~MHz, the present FD approach employs slightly coarser spatial discretization ($c = 0.028$~mm) and lower Courant numbers ($C = 0.2$--$0.7$). Despite the coarser discretization, the present results show lower maximum errors (${\sim}0.08\%$ at $C = 0.2$ versus ${\sim}0.23\%$ at $C = 0.4$ for the FE approach~\cite{Huang_2020}), with both methods exhibiting decreasing phase errors as the Courant number approaches unity.

\subsubsection{Evaluation of numerical amplitude errors}
\label{sec:IV:A:3}
Numerical amplitude errors were evaluated using the same three-dimensional aluminum models (Table~\ref{tab:TABLEI}) and excitation parameters as in the phase error analysis. The relative amplitude error computed from the FFT results is
\begin{align}
\delta_A &= \frac{A(z_2, f)}{A(z_1, f)}-1, 
\label{Eq50}
\end{align}
where $A(z_1, f)$ and $A(z_2, f)$ are the amplitudes of the stress component $\sigma_z$ at two positions $z$ and frequency $f$.

\begin{figure}[t]
\centering
\includegraphics[width=0.8\textwidth]{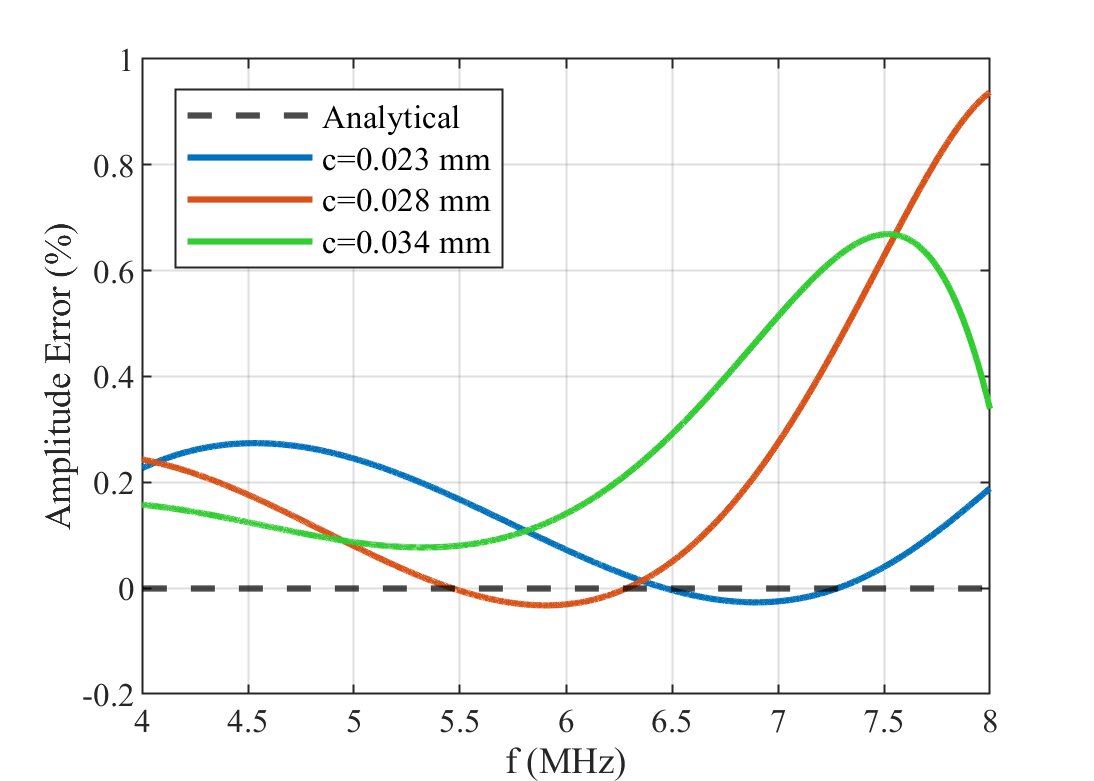}
\caption{Numerical amplitude errors as a function of frequency (4--8~MHz) for the three mesh sizes (0.023, 0.028, and 0.034~mm) at Courant number $C = 0.3$. Each amplitude spectrum was computed using a $2^{20}$-point FFT. The dashed black line at $0\%$ represents ideal behavior.}
\label{fig:FIG6}
\end{figure}

Figure~\ref{fig:FIG6} shows numerical amplitude errors in the range $4$--$8$~MHz for three mesh sizes ($0.023$, $0.028$, and $0.034$~mm) with $C = 0.3$. The $0.023$~mm mesh shows the most stable performance with errors below $0.3\%$, while the $0.034$~mm mesh exhibits the largest variations, peaking at $0.67\%$ near $7.5$~MHz. Finer meshes provide more consistent amplitude accuracy, albeit at higher computational cost.

The Amplitude errors are higher than those reported for the FE methods~\cite{Huang_2020}, which is attributable to fundamental differences in formulation. FE methods preserve energy conservation through element volume integration, whereas the FD point-wise approximation does not explicitly enforce energy conservation between nodes. Nevertheless, amplitude errors remain within acceptable limits ($<1\%$) in the $4$--$8$~MHz range, demonstrating the practical viability of the stress-based FD method for ultrasonic wave propagation simulations.

The stress-based FD implementation demonstrates computational efficiency, completing simulations with more than one billion degrees of freedom (Table~\ref{tab:TABLEI}) in approximately one hour using the dual-GPU setup.

\subsection{Reflection and transmission coefficients}
\label{sec:IV:B}
Material interfaces with strong property contrasts can cause numerical instabilities in FD simulations. Previous studies~\cite{Moczo_2002, Saenger_2004, Bohlen_2006, Kristek_2010} have proposed various solutions, including averaging material properties across discontinuities and employing specialized staggered grids, demonstrating the critical importance of accurate interface treatment in FD schemes.

To evaluate the stress-based formulation at bimaterial interfaces, an FD scheme was implemented that directly solves for stress components across material discontinuities without explicit property averaging. The Stress components are calculated on each side of the interface using the respective material properties, and stress continuity is enforced at the interface. The scheme employs a dual-GPU architecture with parallel computations split across material domains and synchronized data exchange at the interface.

\begin{figure}[t]
\centering
\includegraphics[width=0.8\textwidth]{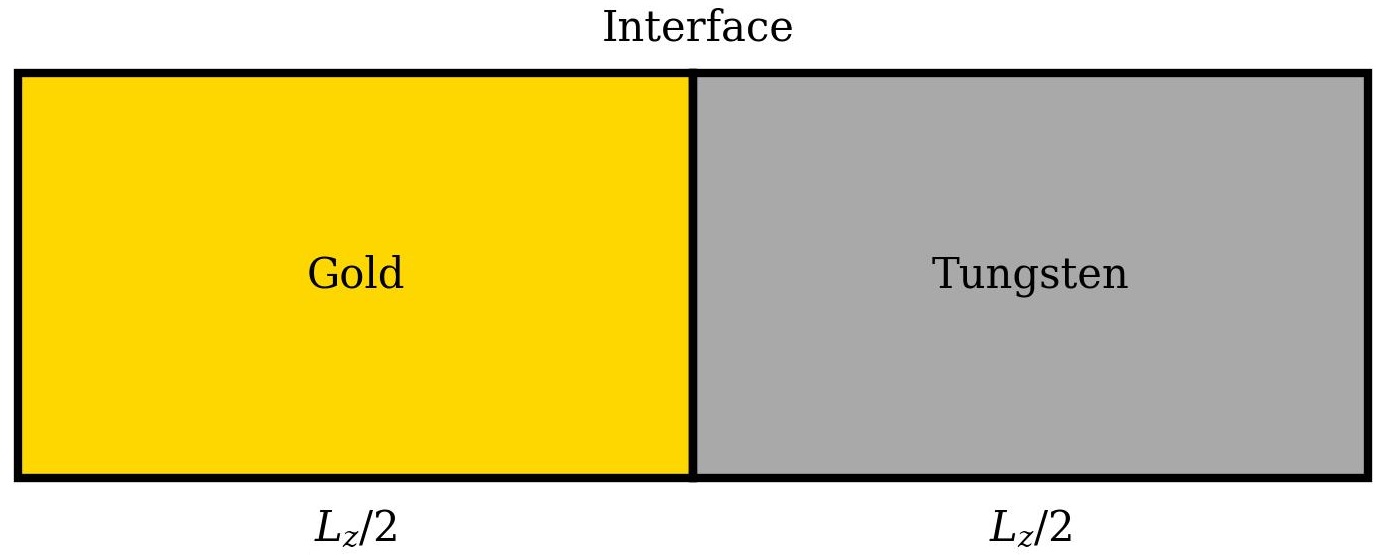}
\caption{Schematic of the bimaterial domain consisting of gold (left) and tungsten (right) separated by a vertical interface. Each material occupies half the domain length ($L_z/2$), with similar densities but distinct elastic properties.}
\label{fig:FIG7}
\end{figure}

This approach was validated by analyzing elastic wave propagation in a gold--tungsten bimaterial domain (Figure~\ref{fig:FIG7}). These materials have similar densities ($\rho_{\text{Au}} = 19{,}300$~kg/m$^3$, $\rho_{\text{W}} = 19{,}250$~kg/m$^3$) but distinct elastic constants ($C_{11} = 213.16$~GPa, $C_{12} = 150.92$~GPa, $C_{44} = 31.12$~GPa for gold; $C_{11} = 523$~GPa, $C_{12} = 203$~GPa, $C_{44} = 160$~GPa for tungsten). To isolate the effects of discontinuities on elastic properties, a uniform density of $\rho = 19{,}300$~kg/m$^3$ was used throughout the simulations.

In this analysis, the second model from Table~\ref{tab:TABLEI} ($401 \times 401 \times 1201$ grid points, $0.028$~mm spacing) was used to isolate the interface effects in the isotropic case. The time step was determined using the CFL condition (Eq.~(\ref{Eq40})) with $C = 0.3$ and the maximum wave speed $V_{\max} = \sqrt{C_{11}/\rho}$ evaluated for both materials, yielding $\Delta t = 1.63$~ns.

Theoretical reflection and transmission coefficients for the stress and displacement formulations are
\begin{align}
R_s = \frac{Z_2-Z_1}{Z_1+Z_2},
T_s=\frac{2Z_2}{Z_1+Z_2},\notag\\
R_d = \frac{Z_1-Z_2}{Z_1+Z_2},
T_d=\frac{2Z_1}{Z_1+Z_2},
\label{Eq51} 
\end{align}
where $Z_1$ and $Z_2$ are the characteristic acoustic impedances of incident (gold) and transmitted (tungsten) media, respectively, and the subscripts $s$ and $d$ denote the stress-based and displacement-based approaches. For the gold--tungsten interface,
\begin{align}
Z_1 &= \sqrt{\rho_{\text{Au}}C_{33_{\text{Au}}}} = 6.0715 \times 10^7~\text{kg/(m}^2\text{s)}
\notag\\
Z_2 &= \sqrt{\rho_{\text{W}}C_{33_{\text{W}}}} = 1.0034 \times 10^8~\text{kg/(m}^2\text{s)},
\label{Eq52} 
\end{align}
yielding
\begin{align}
R_s &= 0.246, \quad T_s = 1.246,\notag\\
R_d &= -0.246, \quad T_d = 0.754.
\label{Eq53}
\end{align}

\begin{figure}[t]
\centering
\includegraphics[width=\textwidth]{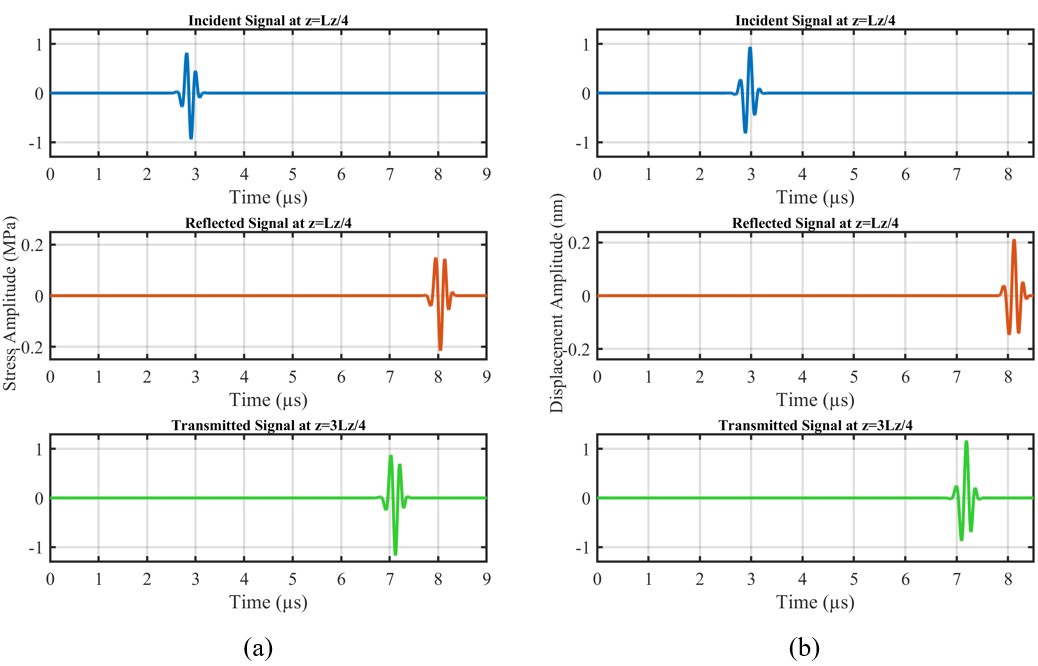}
\caption{Comparison of incident, reflected, and transmitted wave profiles at positions $L_z/4$ and $3L_z/4$ in the gold-tungsten bimaterial domain for (a) stress formulation and (b) displacement formulation. Each panel shows the temporal evolution of waves with amplitude modifications consistent with theoretical reflection and transmission coefficients.}
\label{fig:FIG8}
\end{figure}

These coefficients were calculated from simulations using FFT ratios of incident, reflected, and transmitted signals. For the stress-based formulation, the signed reflection and transmission coefficients are calculated directly from the stress signals and compared to $R_s$ and $T_s$. For the displacement-based formulation, the coefficients are obtained as magnitude ratios of the displacement FFT amplitudes and compared against the theoretical magnitudes $|R_d| = 0.246$ and $T_d = 0.754$.

Figure~\ref{fig:FIG8} compares wave propagation in the gold--tungsten domain for the stress-based and displacement-based formulations, showing incident and reflected waves at $z = L_z/4$ and transmitted waves at $z = 3L_z/4$. The waveforms maintain their characteristic shape with amplitude modifications consistent with the predicted coefficients.

\begin{figure}[t]
\centering
\includegraphics[width=\textwidth]{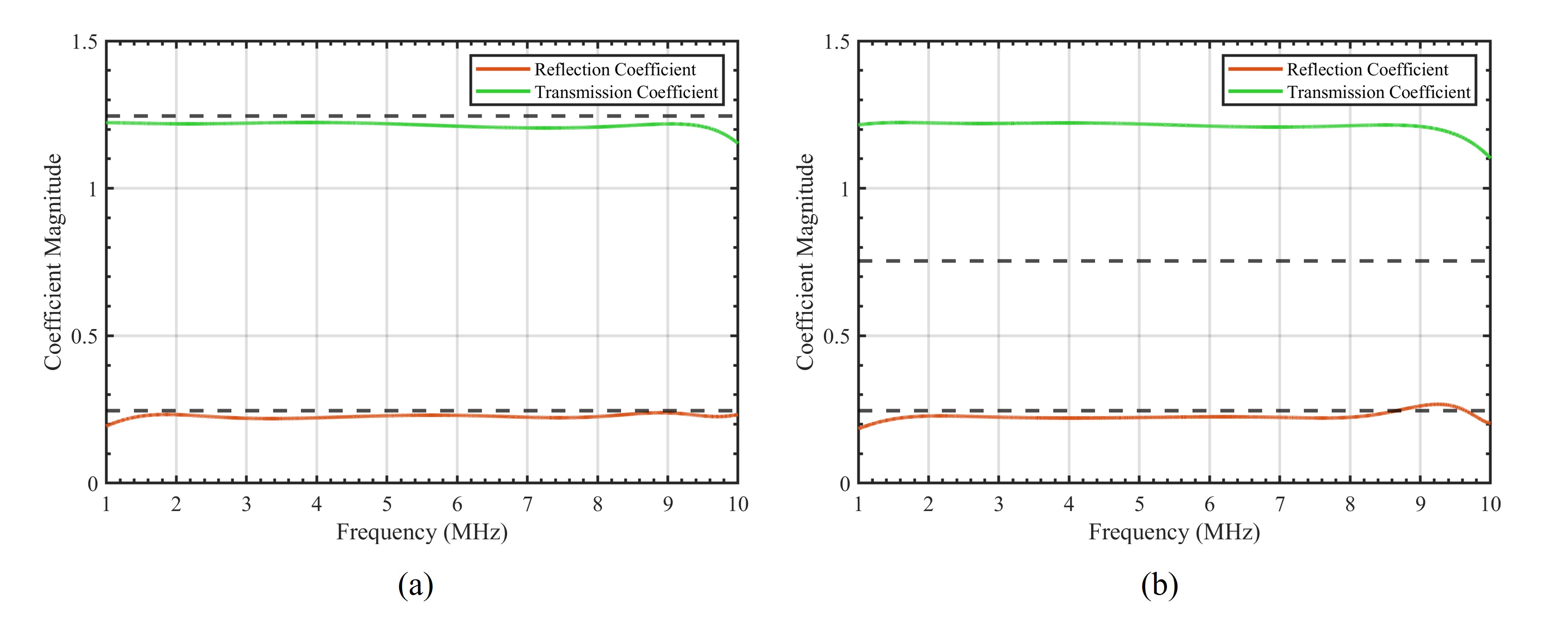}
\caption{Reflection and transmission coefficients as functions of frequency (1--10~MHz) for the gold--tungsten interface comparing (a) stress formulation, and (b) displacement formulation, showing magnitude coefficients $|R_d|$ and $T_d$. Dashed black lines indicate the corresponding theoretical values.}
\label{fig:FIG9}
\end{figure}

Figure~\ref{fig:FIG9} compares numerically computed and theoretical reflection and transmission coefficients over $1$--$10$~MHz for the stress-based and displacement-based formulations. At $5$~MHz, the stress formulation yields $R_s = 0.2287$ and $T_s = 1.2190$ (relative errors of $7.03\%$ and $2.17\%$ with respect to theoretical values of $0.246$ and $1.246$), while the displacement formulation yields magnitude coefficients $|R_d| = 0.2225$ and $T_d = 1.2183$ (relative errors of $9.55\%$ and $61.58\%$ with respect to theoretical magnitudes $|R_d| = 0.246$ and $T_d = 0.754$).

The larger discrepancy in displacement-based transmission coefficients (Figure~\ref{fig:FIG9}b) is due to two factors: (1) numerical approximation of material property derivatives at the interface, where properties change abruptly, and (2) simultaneous satisfaction of displacement continuity and traction conditions. The stress-based formulation (Figure~\ref{fig:FIG9}a) avoids material property derivatives and naturally decouples the boundary equations, which yields better agreement with the theoretical values.

Deviations at frequency extremes arise from FFT limitations: at low frequencies ($<2$~MHz), noise amplification occurs when dividing by small incident signals with fewer cycles, while at high frequencies ($>9$~MHz), discretization effects and numerical dispersion introduce inaccuracies. The close agreement observed in the primary frequency range validates the accuracy of the stress-based formulation at the material interfaces.

\begin{figure}[t]
\centering
\includegraphics[width=0.8\textwidth]{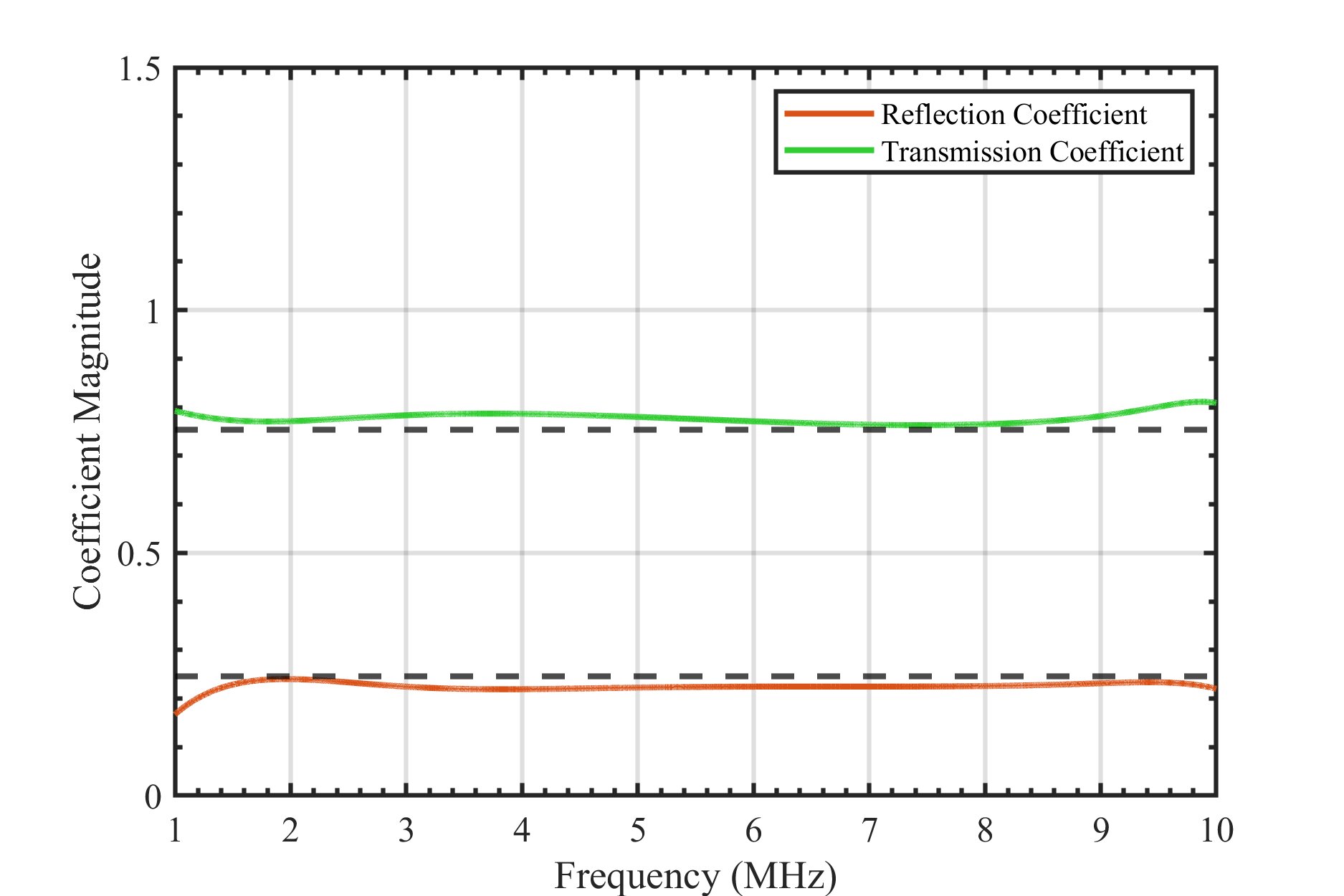}
\caption{Reflection and transmission coefficient magnitudes as functions of frequency (1--10~MHz) for the gold--tungsten interface obtained using Pogo finite element software. Red curve shows $|R|$ and green curve shows $T$, while dashed black lines indicate the theoretical values $|R_d| = 0.246$ and $T_d = 0.754$.}
\label{fig:FIG10}
\end{figure}

Figure~\ref{fig:FIG10} shows the magnitudes of the reflection and transmission coefficients obtained from the Pogo FE software. At $5$~MHz, FE simulations yield $|R| = 0.2226$ and $T = 0.7801$, which differs from the theoretical magnitudes $|R_d| = 0.246$ and $T_d = 0.754$ by $9.51\%$ and $3.46\%$, respectively. Similarly to the FD results, the Pogo software exhibits oscillations at low ($<2$~MHz) and high ($>9$~MHz) frequencies due to FFT limitations and discretization effects, but maintains stable values in the primary range ($2$--$9$~MHz).

\section{Discussion}
\label{sec:V}
The primary advantage of the stress-based formulation, the elimination of material property derivatives, applies specifically to strong-form methods such as FD schemes. Weak formulation methods (e.g., FE) inherently avoid spatial derivatives of material properties through integration by parts, regardless of the chosen variables. This work demonstrates that for GPU-accelerated FD methods, the stress formulation provides a competitive alternative to displacement-based and velocity-stress FD approaches, achieving accuracy comparable to established FE methods while maintaining the characteristic advantages of explicit FD schemes: efficient GPU utilization, minimal computational overhead per time step, and scalability to billion-degree-of-freedom problems.

Dispersion analysis revealed excellent agreement between numerical and analytical phase errors across all mesh sizes and Courant numbers examined (Figures~\ref{fig:FIG4} and~\ref{fig:FIG5}), compared favorably with previous FE results~\cite{Huang_2020}. The results suggest that Courant numbers approaching the CFL stability limit provide an optimal balance between accuracy and computational efficiency. The Amplitude errors were higher than the phase errors, particularly for coarser meshes at higher frequencies (Figure~\ref{fig:FIG6}), attributed to the point-wise nature of the FD approximations, which lack the energy conservation inherent to FE integration. Nevertheless, the errors remained within acceptable limits ($<1\%$) in the frequency range of interest, demonstrating practical viability for ultrasonic simulations.

The stress-based approach offers two key advantages at material interfaces within the FD framework: (1) the elimination of spatial derivatives of elastic constants, which is problematic in displacement-based FD at material discontinuities (Eqs.~(\ref{Eq4})--(\ref{Eq6})), and (2) natural decoupling at boundaries, where stress components directly satisfy traction continuity without auxiliary equations or averaging schemes~\cite{Moczo_2002}. These advantages are clearly demonstrated at the gold--tungsten interface, where the stress-based FD achieves a substantially lower transmission coefficient error compared to the displacement-based FD (Figure~\ref{fig:FIG9}b). Comparison with the Pogo FE software further confirms that the FD stress-based and FE displacement-based approaches achieve comparable accuracy at material interfaces, as both correctly handle wave transmission and reflection physics through different mathematical frameworks, with the FD stress formulation retaining the additional advantage of explicit time-stepping.

The GPU implementation efficiently handles models with more than one billion degrees of freedom. Although the stress formulation requires six field variables compared to three for the displacement formulation, actual memory consumption also depends on storage for material property gradients, intermediate arrays for mixed derivatives, and auxiliary boundary condition arrays. This efficiency is particularly important for heterogeneous media requiring fine spatial discretization to capture microstructural features at grain boundaries and sub-wavelength scales, where conventional FD approaches often face memory constraints in three dimensions.

Several limitations and future directions warrant attention. The current implementation assumes uniform density, eliminating the $\mathbf{M}$ terms in Eq.~(\ref{Eq23}). Although this is a reasonable assumption for polycrystalline metals with uniform density despite varying elastic properties, future work should implement the logarithmic derivative formulation of Eqs.~(\ref{Eq23})--(\ref{Eq24}) for spatially varying density, extending applicability to composite materials and geological applications. Additionally, incorporating nonlinear effects such as acoustoelasticity~\cite{Kube_2022} and validating the approach for complex heterogeneous materials with realistic microstructures (e.g. polycrystalline materials with thousands of grains) represent important next steps.

\newpage

\section*{Acknowledgments}
This material is based on work supported by the National Science Foundation, United States, under Grant Number 2225215. Any opinions, findings, conclusions, or recommendations expressed in this material are those of the author(s) and do not necessarily reflect the views of the National Science Foundation, United States.

\section*{Author Declarations}

\subsection*{Conflict of Interest}
The authors have no conflicts of interest to disclose.

\section*{Data Availability}
Data supporting the findings of this study are available from the corresponding author on a reasonable request.

\newpage
\bibliographystyle{unsrtnat}
\bibliography{References}

\end{document}